\documentclass[12pt,A4paper]{article} 
\usepackage{indentfirst,bm,geometry,amsmath} 
\usepackage{graphicx}
\usepackage{float}
\usepackage{caption} 
\usepackage{tikz}
\usepackage{enumitem}
\usepackage{booktabs,setspace} 
\usepackage{threeparttable} 
\usepackage{multirow} 
\usepackage{diagbox} 
\usepackage{array} 
\usepackage{rotating} 
\usepackage{amssymb}
\usepackage{hyperref} 
\usepackage{xurl}    %
\usepackage{url}     %
\usepackage[linewidth=1pt]{mdframed}
\usepackage[round,authoryear]{natbib}
\usepackage{amsthm}
\newtheorem{result}{Result}

\hypersetup{         
    colorlinks=true, 
    linkcolor=blue,  
    filecolor=magenta, 
    urlcolor=cyan,   
    citecolor=blue  
}

\title{Do people rely on ChatGPT more than their peers to detect deepfake news?\footnote{This research has benefited from the financial support of (a) the Joint Usage/Research Center, the Institute of Social and Economic Research (ISER), and the University of Osaka; (b) Grants-in-Aid for Scientific Research (Nos. 20H05631, 23H00055, and 25H00388) KAKENHI from the Japan Society for the Promotion of Science; and (c) the Support for Pioneering Research Initiated by the Next Generation program of the Japan Science and Technology Agency (No. JPMJSP2138). The design of the experiment reported in this paper was approved by the IRB of ISER (\#20231001) in October 2023, and the experiment is preregistered at aspredicted.org (\#149838). We thank Tiffany Tsz Kwan Tse and Taisuke Imai for their helpful comments and suggestions. We are also grateful to Bor Hodošček, Takahiro Honda, Hiroshi Mito, Masao Ochi, Yushi Sugimoto, Eri Tanaka, and Yoko Yumoto for providing advice materials in the additional experiment, and to those experts who wished to remain anonymous. Finally, we gratefully acknowledge the support of Yuta Shimodaira, Zhelin Zhang, and Satsuki Yamada in conducting the experiment.}}

\author{Yuhao Fu\thanks{Graduate School of Economics, University of Osaka. E-mail: u889037j@ecs.osaka-u.ac.jp} \and Nobuyuki Hanaki\thanks{Corresponding author. Institute of Social and Economic Research, University of Osaka, and University of Limassol. E-mail: nobuyuki.hanaki@iser.osaka-u.ac.jp}}

\begin{document} 

\maketitle
\begin{abstract}
This experimental study investigates how people rely on different sources of advice when detecting AI-generated fake news (deepfake news). In a laboratory deepfake detection task, student participants identified the proportion of human-written (non-AI-generated) content in synthetic deepfake news articles and received advice from ChatGPT (GPT-4), human peers, or linguistic experts. The results show that participants rely more on ChatGPT than on human peers when detecting GPT-2-generated deepfake news. Participants also rely more on linguistic experts than on peers, while the relative reliance on experts versus ChatGPT is mixed across experimental waves, potentially reflecting time trends in beliefs about AI-based detection. Importantly, in the additional experiment conducted in 2025 under the same experimental procedure, participants relied more on linguistic experts than on ChatGPT. Moreover, performance improvements reflect the joint role of reliance and advice quality, arising primarily when participants rely on high-quality advice. Overall, relying on AI to detect AI-generated deepfakes can improve detection outcomes, but only when AI-based detection tools are of sufficiently high quality. These findings highlight the dual role of GAI as both a source of deepfakes and a tool for mitigating related risks.
\end{abstract}

\hspace{0.4cm}\textbf{Keywords:}  ChatGPT, AI reliance, deepfake detection, advice taking, human--AI interaction 

\medskip

\hspace{0.4cm}\textbf{JEL:} C90; D83; D90; D91 

\newpage
\doublespacing

\section{Introduction}
\noindent Generative AI (GAI) products, such as ChatGPT, have attracted widespread attention since 2022 and have been rapidly adopted across many domains. Alongside this diffusion, growing concerns have emerged regarding their societal harms, particularly the spread of AI-generated misinformation (deepfakes) \citep{lundberg2024potential,sophiA2025social}. In response, recent studies and policy discussions increasingly propose leveraging GAI itself---such as commercial or platform-integrated AI detectors (e.g., Turnitin)---to identify and mitigate deepfake content \citep{patil2024novel,bhattacharjee2024fighting}. However, these tools are imperfect and can misclassify human-written text as AI-generated, producing false positives with tangible real-world consequences \citep{weber2023testing,knilans2024darkside,zhang2024detection}. GAI thus increasingly plays a dual role: \textit{it is both a source of deepfakes and an imperfect tool used to combat them.}

This dual role highlights the importance of understanding individuals’ reliance on GAI---based tools to detect deepfake content---whether they choose to ``fight fire (AI) with fire (AI),'' and whether they appropriately account for both the benefits and risks of doing so. At the same time, such reliance may itself become problematic, as concerns about over-reliance on AI systems have grown substantially \citep{schemmer2023appropriate,klingbeil2024trust}.

Building on this context, this study examines individuals' ``reliance on AI to detect AI". Specifically, we address the following research question:
\begin{enumerate}[label=\textbf{RQ:}, leftmargin=3em]
\item \textit{Do people rely on ChatGPT more than on their human peers to detect deepfake news?}
\end{enumerate}

We conducted a laboratory experiment in which participants performed a deepfake detection task. Specifically, they were asked to identify the proportion of non–AI-generated (human-written) content in synthetic deepfake news articles composed of both human-written and AI-generated text. After providing an initial identification, participants in our main experiments received advice in the form of corresponding identifications generated either by ChatGPT (GPT-4) or by human peers who participated in the same experimental session.

We adopt the ``weight of advice" (WOA) measure \citep{harvey1997taking,yaniv1997precision} to quantify the reliance and find that participants rely more on ChatGPT than on human peers when detecting deepfake news. Moreover, advice from ChatGPT leads to higher performance, reflecting its higher advice quality. Importantly, variation in the proportion of human-written (non-AI-generated) content within articles does not systematically affect either reliance behavior or performance.

We also conducted an additional experiment that introduced two new advice sources---\textit{linguistic experts} and \textit{human peers from earlier experimental sessions}---while also replicating the condition in which participants received advice from ChatGPT but varying the timing of the questionnaire measuring participants' prior beliefs, moving it from after the main task to before the task, to test whether questionnaire placement affects reported beliefs. The results show that questionnaire timing does not significantly affect prior beliefs, and that participants' reliance on human peers does not differ depending on whether peers participated on the same day or in earlier sessions. Participants rely more on linguistic experts than on human peers, while the relative reliance on experts compared to AI is mixed---lower than reliance on AI in the main experiment but higher in the additional experiment. We interpret this pattern as reflecting time trends in beliefs about AI-based deepfake detection.

In addition, we examine how participants' prior beliefs---capturing their subjective preferences over advice sources---and advice quality jointly shape performance improvement. Our results indicate that improvements in detection accuracy arise from the interaction between advice quality and reliance: participants rely more on sources they personally trust, and the benefits of high-quality advice materialize only when advice is actually used.

Because the WOA measure does not totally utilize all observations---and also as a robustness check for the main analysis---we further adopt the two-stage framework of \citet{vodrahalli2022humans} and implement it using a Heckman selection correction model \citep{heckman1974shadow,heckman1979sample}. This approach decomposes reliance into an “activation” stage (whether advice is taken) and an “integration” stage (how strongly it is incorporated). The results validate this decomposition: advice source and prior beliefs systematically shape both stages, while the advice–initial gap (gap between the advice and initial response) exerts opposite effects---encouraging activation but dampening integration.

These findings have direct implications for policy frameworks aimed at mitigating the risks of deepfake misinformation. They suggest that, while people tend to trust AI-based detection tools, the effective use of such tools depends not only on public beliefs and trust but also on their objective detection quality. Accordingly, regulators and institutions may consider promoting the development and responsible deployment of AI-based detection systems, while recognizing that policy outcomes are jointly shaped by detectors' performance and users' beliefs. In this sense, encouraging the adoption of high-quality AI-based detection tools may be a more effective policy priority than relying on source credibility or trust alone, without regard to realized detection performance.

The remainder of this paper is organized as follows. \textbf{Section}~\ref{section2} reviews previous studies on deepfake detection and AI reliance. \textbf{Section}~\ref{section3} presents the experimental design and hypotheses. \textbf{Section}~\ref{section4} summarizes the results of the main experiments. \textbf{Section}~\ref{section5} reports the additional experiments. \textbf{Section}~\ref{section6} provides the discussions of the findings, and \textbf{Section}~\ref{section7} concludes the paper.

\section{Literature Review}
\label{section2}

\noindent We review two strands of research. First, we summarize the literature on deepfake detection, focusing on its societal relevance and detection challenges. Second, we review prior work on reliance on AI advice, with attention to behavioral mechanisms, measurement approaches, and applications across domains.

\subsection{Background on Deepfake Detection}

\subsubsection{Why Deepfakes Matter}

\noindent Deepfakes refer to synthetic media---primarily audio and video---generated using deep learning techniques \citep{chadha2021deepfake}. These outputs are designed to closely mimic real content, which makes them potentially harmful to society \citep{katarya2020study, sareen2022threats}. With the rapid advancement of GAI, the concept of deepfakes has expanded beyond audio and video to include textual content \citep{chong2023bot, uchendu2023does, uchendu2024catch}, making the creation and spread of deepfakes increasingly easy and widespread.

Deepfakes pose serious risks across multiple domains. In politics, they can increase cognitive load, reinforce confirmation bias, and undermine social trust, posing particular threats to elections and democratic processes \citep{islam2024ai, amin2025influence, gupta2025influence}. In academia and education, the use of deepfake data and images creates systemic risks to research integrity \citep{chen2024research, chauhan2024impact}. Deepfakes also threaten individuals' daily lives by enabling fraud, identity misuse, and reputational harm, such as deepfake pornography created using others' images \citep{umbach2024non} and voice-based scams using synthetic speech \citep{barrington2025people}. 

In response, scholars emphasize the need for coordinated efforts across policy, technology, and human decision-making: governments are urged to develop targeted legal responses \citep{yamaoka2019disrupting, ramluckan2024deepfakes}, researchers continue to improve detection methods and tools \citep{mirsky2021creation}, and users increasingly seek ways to verify media in everyday social contexts as deepfakes become harder to distinguish from real content \citep{ahmed2023perception}. Complementing detection-based approaches, recent work also explores human-centered interventions that aim to mitigate deepfake harms by shaping beliefs and source perceptions prior to exposure, such as pre-emptive source discreditation (i.e., warning users about the unreliability of a source before exposure) and debunking of AI-generated misinformation \citep{spearing2025countering}.

\subsubsection{Human and Machine Detection of Deepfakes}
\noindent Growing evidence shows that humans perform poorly when trying to detect deepfake content. \citet{chen2023can} find that AI-generated misinformation---such as outputs from GPT-3.5---is harder for people to identify than human-written fake news, making it potentially more harmful. A systematic review and meta-analysis by \citet{diel2024human} reaches a similar conclusion: across 56 studies, average detection accuracy is only 55.54\% (and just 52\% for deepfake text). Consistent with these findings, \citet{groh2024human} report that increasing the share of deepfakes in a set of political speeches does not meaningfully change people's judgments, and that deepfake text is even harder to detect than manipulated audio or video.

Compared with human detection, much of the recent literature focuses on machine-based approaches to deepfake detection. As \citet{zellers2019grover} argue, ``the best way to detect neural fake news is to use a model that is also a generator.'' Empirical studies provide partial support for this view. For example, \citet{alexander2024can} show that GPT-4 can identify misleading visualizations with moderate accuracy even without task-specific training, and \citet{koka2024evaluating} report detection accuracy exceeding 95\% for GPT-4 on deepfake news. However, detection performance is far from uniform. \citet{sallami2024deception} find that while GPT-4 performs well on AI-generated misinformation, it is substantially less accurate when detecting human-created fake news. Public AI detectors exhibit even greater instability: although some benchmark studies report high accuracy rates \citep{koka2024evaluating,sallami2024deception,liu2024great}, others document performance only slightly above chance. In particular, \citet{weber2023testing} show that widely used detectors---including commercial systems such as Turnitin---are easily fooled by paraphrasing, a conclusion echoed by the review in \citet{chaka2024reviewing}. As GAI continue to advance, both deepfake generation and detection technologies evolve rapidly, creating an ongoing ``generation--detection'' arms race \citep{laurier2024cat}.

Importantly, these performance limitations are not merely technical but can translate into real-world risks. A growing body of evidence documents that AI detectors frequently misclassify human-written text as AI-generated, leading to false positives \citep{weber2023testing,knilans2024darkside,zhang2024detection}. Even OpenAI cautions that AI-writing detectors are not reliable for high-stakes judgments.\footnote{OpenAI Help Center: ``\href{https://help.openai.com/en/articles/8313351-how-can-educators-respond-to-students-presenting-ai-generated-content-as-their-own}{Do AI detectors work? In short, not in our experience.}''} Well-known cases include detectors labeling the \emph{U.S. Constitution} and passages from the \emph{Bible} as AI-generated,\footnote{See Ars Technica on the Constitution case \citep{Edwards2023ArsDetectorsDontWork} and India Today on Bible passages being flagged as AI \citep{Chakravarti2023BibleClassifier}.} and false positives have led to tangible harms in education, such as publicized misconduct cases and delayed graduations \citep{ABC2025SectorWide,Spectrum2025UB,AdelaideNow2025ACU,CourierMail2024AI}.

Beyond individual performance, collaboration has been shown to substantially enhance human deepfake detection. \textit{Human–Human collaboration} improves accuracy: \citet{uchendu2023does} find that group discussion leads to better detection of deepfake text than individual judgment for both experts and non-experts, and \citet{groh2022deepfake} show that aggregating human predictions on deepfake videos yields accuracy comparable to state-of-the-art detectors and clearly above that of single raters. These effects align with the review in \citet{diel2024human}. \textit{Collaboration with AI} appears similarly promising: Experiments in \citet{groh2022deepfake} show that participants who observe an AI model's prediction outperform both standalone humans and the model itself. \citet{diel2024human} likewise note systematic accuracy gains when humans receive AI assistance, and \citet{somoray2025human} demonstrate that humans and models rely on different cues when judging authenticity---suggesting complementarities that human--AI collaboration can leverage.

\subsubsection{Experimental Innovations}

\noindent While deepfakes pose serious societal risks and collaboration with AI or other humans can improve detection accuracy, neither human nor machine detection is perfect. As a result, uncritical reliance on external detection results may introduce secondary risks, particularly in high-stakes contexts. This tension motivates our examination of how individuals rely on AI tools versus human collaboration when detecting deepfake news.

In this study, we focus on a common form of deepfake---deepfake news---and examine a deepfake detection task in a laboratory setting in which participants identify the proportion of non-AI-generated text in each article. Although this design cannot fully replicate real-world environments, it provides an initial and controlled way to measure how individuals perceive deepfake content and how they behave when detecting it. 

Importantly, the task is designed to capture detection of AI-generated content embedded in deepfake news articles, rather than the identification of human-written fake news per se. Most experimental studies in economics on human-written fake news detection ask participants for a binary judgment (real vs.\ fake) \citep{serra2021mistakes,arin2023ability,thaler2024fake}. By contrast, we ask participants to report the article's non-AI-generated proportion (human-written proportion), following recent work that moves beyond whole-document labels toward partial detection and localization \citep{zeng2024detecting,zhang2024machine}. We use a proportion for four reasons.
First, our stimuli include both totally AI-generated and totally human-written items, so a proportion nests binary judgments while offering finer measurement.
Second, as models improve, binary human detection becomes unreliable, whereas a proportion better captures subjective uncertainty.
Third, real-world content often mixes AI output with human editing, making proportion ratings more aligned with actual production processes.
Fourth, modern AI detectors themselves output continuous scores or estimated “AI-generated proportions,” rather than hard labels.

We incorporate both human–human and human–AI collaboration. Participants receive advice either from peers' initial identifications or from identifications generated by ChatGPT. This design enables us, for the first time in the deepfake-detection context, to directly compare how individuals rely on and respond to these two distinct modes of collaboration.

We also introduce a third source of advice: linguistic experts. Because the development of GAI---especially large language models (LLMs)---relies heavily on linguistic knowledge and analysis, linguists and linguistically trained annotators are frequently involved in dataset curation, model evaluation, and assessments of whether LLM-generated text resembles human-written language \citep{bender2021dangers, ouyang2022training, workshop2022bloom}. Introducing this additional advice source allows us to investigate a previously unexplored question: \textit{do people trust experts who are familiar with, and in some cases involved in, the development of LLMs to accurately identify AI-generated text}? This aspect of human judgment has not been examined in existing deepfake detection or experimental economic studies.

\subsection{Previous Work on AI Reliance}
\subsubsection{Algorithm Aversion, Appreciation, and Human Perception of AI}
\noindent
A large body of research examines how individuals respond to algorithmic advice relative to human judgment. A central finding is that people do not treat algorithmic and human advice symmetrically: reliance on algorithms varies systematically across contexts, ranging from algorithm aversion, where individuals rely less on algorithms than on humans \citep{dietvorst2015algorithm}, to algorithm appreciation, where algorithmic judgments receive greater weight \citep{logg2019algorithm}.

Evidence of ``algorithm aversion" has been documented across a range of settings. For example, experts often rely less on algorithmic systems than on non-expert humans \citep{reverberi2022experimental, agarwal2023combining}, and similar patterns arise when algorithms are compared with human peers rather than experts \citep{gaube2021ai, mesbah2021whose}. In contrast, \citet{logg2019algorithm} show that individuals may rely more on algorithms than on humans when algorithms are perceived as appropriate for the task. Related work also finds that allowing users to make small adjustments to algorithmic advice can reduce algorithm aversion and increase reliance \citep{dietvorst2018overcoming}. These mixed findings indicate that algorithm aversion is not a fixed bias but depends on how algorithms are perceived and used.

Subsequent research explores the mechanisms underlying these divergent responses. \citet{castelo2019task} show that reliance on algorithms depends on task characteristics, particularly perceived objectivity: algorithm aversion is stronger in subjective tasks than in objective ones. They further find that increasing the perceived human likeness of algorithms can mitigate aversion in subjective domains. Other studies emphasize the role of interaction and control. For instance, \citet{maggioni2023if} show that verbal interaction with robots reduces algorithm aversion, while \citet{tse2024beware} find that granting decision makers greater freedom in final decisions can induce over-reliance on algorithms, even when algorithmic performance is low. Finally, focusing on process design in high-stakes settings, \citet{yin2025designing} examine how the timing of AI advice affects diagnostic decision making and find that physicians perform best when AI advice is provided after an initial diagnosis, and worst when no AI advice is available.

Recent work further synthesizes these findings by focusing on how individuals conceptualize and evaluate AI itself. As AI technologies have evolved, the notion of AI has expanded beyond task-specific algorithms to encompass a broad set of interrelated technologies, including algorithms, decision-support systems, social robots, and conversational agents such as ChatGPT \citep{walsh2019effective, baines2024advice}. In particular, \citet{baines2024advice} review a wide literature on advice from AI across domains, emphasizing how trust, acceptance, and reliance on AI advice depend on contextual and individual factors. Complementing this perspective, \citet{passi2025addressing} synthesize evidence from over 120 interdisciplinary studies to examine the negative consequences of AI mistakes, with particular attention to overreliance on AI---especially GAI---and the antecedents and mitigation strategies associated with such overreliance. Similarly, \citet{chevrier2025forth} propose a conceptual framework that organizes prior evidence around key dimensions such as perceived competence, accountability, and controllability in the advice-taking process, especially for ChatGPT. Consequently, this integrative body of work suggests that insights from earlier studies on human-algorithm interaction can be naturally extended to contemporary research on human--AI interaction, including reliance on GAI systems.

\subsubsection{Applications of AI Reliance Across Domains}
\noindent
More recently, research on reliance on AI advice has expanded to a broader set of application domains. In economic and decision-making settings, recent work examines how individuals rely on AI advice across diverse contexts, including strategic games \citep{klingbeil2024trust}, financial forecasting and investment decisions \citep{kim2024trust}, organizational and managerial decision making \citep{stiefenhofer2026delegating}, and labor-related tasks shaped by GAI \citep{brynjolfsson2025generative}.

Beyond economics, medical contexts constitute a major area of study on reliance on AI advice across multiple stages of clinical decision making. Existing work examines how AI advice affects diagnostic processes and information integration \citep{yin2025designing}, as well as trust, acceptance, and explainability of medical AI systems \citep{mainz2024medical, kuper2025psychological, rosenbacke2024explainable, tun2025trust}. A growing literature also focuses on patients' perceptions of AI- and ChatGPT-based healthcare advice, documenting how trust and acceptance vary across tasks and informational settings \citep{chen2024perceptions, sun2024trusting, van2024if, chen2025impact, kelly2025factors}.

Emerging evidence from other domains, including law \citep{tamo2024regulating} and education \citep{viberg2025explains, amoozadeh2024trust}, further suggests that reliance on AI advice is a domain-general phenomenon shaped by institutional context and task structure.

\subsubsection{Measuring AI Reliance}

\noindent The judge–advisor paradigm (JAS) \citep{sniezek1989social} is widely used in economics and psychology to study the reliance on algorithm. In this framework, the reliance is commonly quantified using the WOA, which measures the extent to which individuals adjust their initial judgments toward the advice. A comprehensive meta-analysis by \citet{bailey2023meta} shows that individuals, on average, place substantially less than equal weight on advice and that WOA varies systematically with task characteristics, advisor attributes, and decision makers' confidence, underscoring that WOA captures a behavioral response rather than a stable preference parameter. In studies of algorithm aversion, WOA has been widely used to quantify reliance on algorithmic advice \citep{dietvorst2015algorithm, dietvorst2018overcoming, logg2019algorithm}, and more recent work extends this approach to measure reliance on GAI, including LLMs such as ChatGPT \citep{zhang2023taking,rebholz2024enhancing, boob2025ai, bo2025rely}.

Beyond behavioral advice-taking measures, a large literature studies responses to AI advice using survey-based approaches. Many studies adopt acceptance-oriented frameworks such as the Technology Acceptance Model (TAM), which focus on perceived usefulness, ease of use, and intentions to adopt AI systems \citep{davis1989perceived, venkatesh2003user}. For example, \citet{biswas2024influence} examine how educational background and self-perceived technological proficiency relate to reliance on AI, and \citet{setyaningsih2025efl} study students' reliance on ChatGPT's writing suggestions within a TAM-based framework. Other survey-based studies applied direct elicitation, asking individuals to report whether, or to what extent, they rely on AI advice, sometimes using vignette-based designs \citep{hoff2015trust, bussone2015role, glikson2020human, rosenbacke2024explainable}. Such approaches are widely used in applied domains, including medicine \citep{bussone2015role}, law \citep{eckhardt2025survey}, and management \citep{glikson2020human}. While informative about attitudes, intentions, and stated use of AI advice, these survey-based measures do not directly capture how advice is integrated into incentivized decision making.

\subsubsection{Benchmarking AI Advice: Human Peers and Experts}
\noindent
Across these approaches, reliance on AI advice is typically evaluated relative to benchmark sources of judgment, most commonly human peers or domain experts. These benchmarks serve different purposes and correspond to distinct research questions.

Human peers provide a natural and methodologically conservative benchmark for evaluating reliance on AI advice. Because peer judgments reflect lay people's decision making, they allow researchers to isolate the effect of advice source without conflating it with differences in expertise or authority. Using this benchmark, prior studies compare reliance on AI advice with reliance on advice from non-expert humans holding objective quality constant. For example, \citet{vodrahalli2022humans} show that individuals are more likely to activate advice labeled as AI than otherwise identical advice attributed to human peers. Similarly, \citet{zheng2025students} compare reliance on AI advice and human peer advice in numerical estimation tasks and find no systematic tendency toward greater reliance on AI than on peers. Despite these contributions, direct comparisons between human peers and GAI systems---particularly ChatGPT---remain relatively limited, leaving open questions about how reliance on GAI differs from reliance on ordinary human judgment.

In contrast, domain experts provide a normative benchmark, especially in high-stakes or specialized tasks where decision quality rather than conformity to typical human judgment is the primary concern. A growing literature compares reliance on AI advice with reliance on expert advice across domains. For instance, \citet{agrawal2023oecd} examine reliance on AI versus human experts across OECD and Indian samples, finding that participants evaluate experts and AI differently along dimensions of competence, trust, and moral responsibility. \citet{larkin2022paging} study reliance on AI and human expert advice in medicine and finance and show that participants update their decisions more in response to expert recommendations, although the magnitude of this effect varies by context. However, direct evidence comparing reliance on GAI systems such as ChatGPT with reliance on human experts remains scarce, particularly in controlled experimental settings that allow for clean measurement of advice integration.

\subsubsection{Contributions}
\noindent
Relative to the existing literature on AI reliance, this study makes three main contributions.

First, we study reliance on AI advice in the context of deepfake detection, a setting that has become increasingly salient with the rapid improvement of GAI technologies. Our design allows us to examine how individuals rely on AI assistance when evaluating content generated by AI itself, highlighting the dual role of GAI. In light of the rapid spread of deepfakes and the growing use of AI-based detection tools in real-world contexts, our analysis helps inform more effective approaches to the governance and regulation of deepfake content.

Second, we directly compare advice from ChatGPT with advice from human peers and domain experts within a unified experimental framework. Although recent studies examine reliance on AI advice in isolation or relative to a single benchmark, systematic comparisons across AI, peers, and experts remain scarce, particularly in controlled settings that allow clean identification of source effects.

Third, we adopt the JAS paradigm using the WOA metric and further examine the advice-taking process by decomposing reliance using activation-integration model, providing additional evidence on the mechanisms through which AI advice influences decision making.

\section{Experimental Design}
\label{section3}

\subsection{Procedure}
\begin{figure}[t]
\centering
\begin{tabular}{c}
(a) Overall Procedure\\
\includegraphics[scale=0.8]{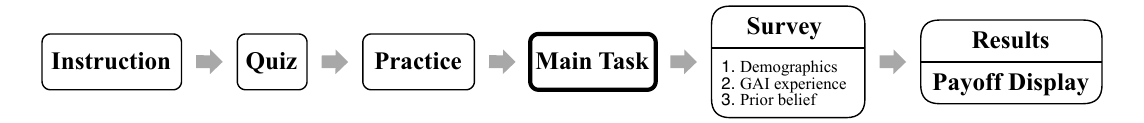}\\
(b) Main Task\\
 \includegraphics[scale=1.15]{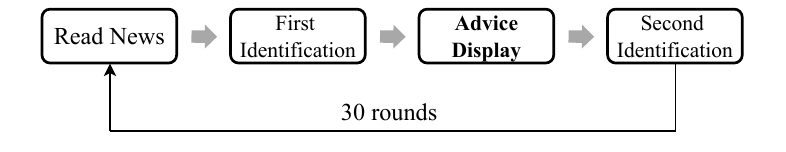}\\
\end{tabular}
\caption{Experimental Procedure: (a) Overall Procedure and (b) Main Task}\label{design}
\end{figure}

\noindent The experiment was programmed using oTree 5 \citep{chen2016otree}, and the overall procedure is shown in Panel (a) of Figure~\ref{design}.

In the experiment, after reading through the instructions (See an English translation in Online Appendix F), each participant was asked to take a quiz (see Online Appendix G) to ensure they understood the rules. Then, they practiced once and entered the main tasks. After finishing all the tasks, they were asked to complete some survey questions, and the results and the final payoff were shown.

\subsection{Main Task}
\noindent The main task consists of 30 rounds of a deepfake detection task implemented within the JAS framework, as illustrated in Panel (b) of Figure~\ref{design}. Representative experimental screens are provided in Online Appendix~I.1.

There are four stages in each round. Participants first read a deepfake news article and report their initial identification of the proportion of human-written content in the article ($HMpro$). They then receive advice from ChatGPT or Human peers. After receiving the advice, participants are asked to submit a second identification. No time constraints are imposed, except for a 10-second Advice Display stage. 

Details of the deepfake news materials, the deepfake detection task, treatments and two identifications are described below.

\subsubsection{Deepfake News Materials and Deepfake Detection}
\noindent The deepfake news articles were Japanese deepfake news collected from an open deepfake news dataset.\footnote{\href{https://github.com/tanreinama/japanese-fakenews-dataset?tab=readme-ov-file}{https://github.com/tanreinama/japanese-fakenews-dataset?tab=readme-ov-file}} We randomly selected 30 news articles, primarily covering topics such as politics, sports, meteorology, and public safety.\footnote{The original Japanese texts are available upon request. Detailed category definitions and short descriptions of the news content are provided in Online Appendix~E.3.} The news items fall into three types, as summarized in Table~\ref{news_category_short}, where ``Length'' refers to the number of characters.

\begin{table}[t]
\centering
\begin{threeparttable}
\caption{News Materials} \label{news_category_short} 
\begin{tabular}{lcccc}
\toprule
Type            & Count & Min. Length & Max. Length & \(HMpro\) \\
\midrule
Totally real    & 10    & 317         & 460         & 100 \\
Totally fake    & 10    & 309         & 462         & 0 \\
Partially fake  & 10    & 323         & 393         & $(0,100)$ \\
\bottomrule
\end{tabular}
\end{threeparttable}
\end{table} 

The totally real news articles were written by humans and collected from Japanese
Wikinews,\footnote{\href{https://ja.wikinews.org/wiki}{https://ja.wikinews.org/wiki}} the totally fake news articles were
generated by OpenAI's Japanese GPT-2 model, and the partially fake news articles
consisted of both real and fake content. In the partially fake news, the first
part of the article was human-written and the second part was AI-generated; the human-written part always appeared first.

\begin{figure}[t]
\centering
\fbox{\includegraphics[scale=0.32]{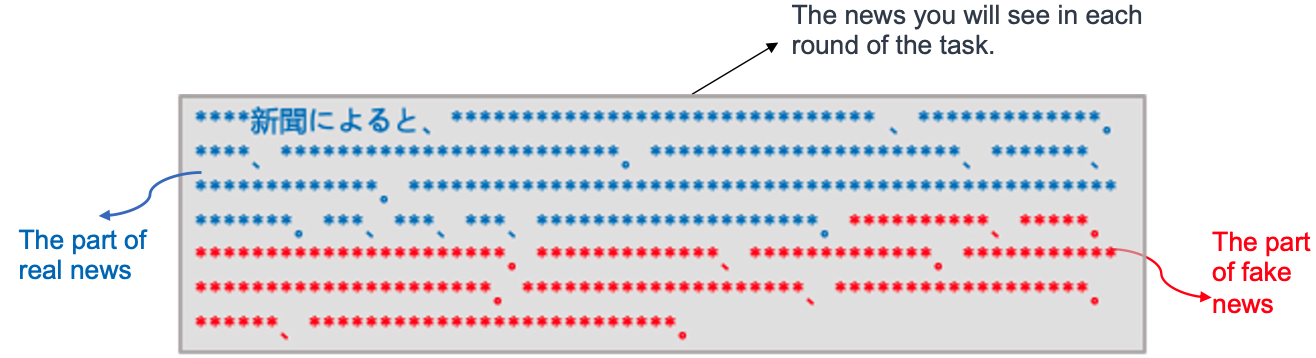}}
\caption{Instructional Illustration of News Composition Shown to Participants}\label{Instructional_Illustration}
\end{figure}

Figure~\ref{Instructional_Illustration} illustrates the explanatory guidance provided to participants in the experimental instructions regarding the composition of news articles. The proportion of the ``real'' (human-written) part, denoted as $HMpro$, is defined as$$\begin{aligned}HMpro_r=\frac{\text{the length of human-written part of the News in Round }r}{\text{the length of the News in Round }r} \times100\end{aligned}\text{,}$$ where $HMpro_r=0$ represents totally fake news, $HMpro_r=100$ represents totally real news, and $HMpro_r\in(0,100)$ represents partially fake news in round $r$.

The \textbf{deepfake detection task} in this study is therefore defined as follows: \textit{participants read a deepfake news article and identify its} $HMpro$ (the proportion of the ``real'' (human-written) part), which is operationally interpreted as the degree of “authenticity.” In this context, “authenticity” serves purely as an operational label for $HMpro$ and does not refer to the factual truthfulness of the news content.

\begin{figure}[t]
\centering
\fbox{\includegraphics[scale=0.67]{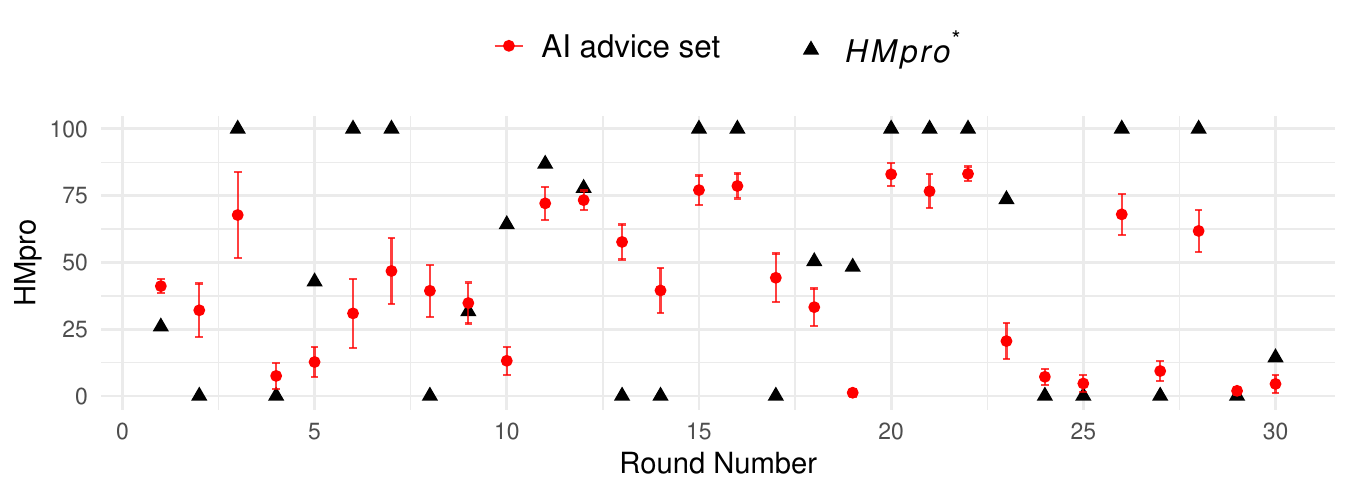}}
\caption{$HMpro^*$, the original AI advice set and Round Number}\label{sequence_HMpro}
\caption*{Note: The points marked with ``$\blacktriangle$'' represent the true $HMpro$ values in the 30-round tasks ($HMpro^*$). The red points indicate the original AI advice (24 data points in each round) generated before the experiment, and the 95\% CI.}
\end{figure}

In the instructions, participants are clearly informed about the composition style of the news materials, along with the definition of $HMpro$. They are also informed that $HMpro$ represents the degree of ``authenticity'' they must identify and report. The true $HMpro$ of 30 pieces of news ($HMpro^*$) were assigned in a random sequence as shown in Figure~\ref{sequence_HMpro}.

\subsubsection{First Identification}

\noindent After each participant had read the news, they were asked to report a number between 0 and 100 to represent their first identification of $HMpro$ with a slider (see Figure~I.2 in Online Appendix~I.1). Each participant was required to submit their first identification, $Initial\ Response$, in this stage; otherwise, they could not proceed to the next page.

\subsubsection{Treatments by Advice Source: AI vs.\ Human}
\noindent
The Advice Display Stage (see Figures~I.3--I.6 in Online Appendix~I.1) follows a \textit{between-subject} design with two treatments that differ only in the source of advice: an \textbf{AI} treatment and a \textbf{Human} treatment. The details are as follows,

\vspace{1em}
\noindent \textbf{AI treatment.} The advice was one response randomly selected from 24 responses generated by ChatGPT using the GPT-4 model before the experiment, using the following prompt:

\begin{center}
\noindent\fbox{
    \parbox{.9\linewidth}{\textit{-- We will now send you some Japanese news. Please identify how real it is and report your belief in its authenticity as an integer from 0 to 100, with 0 representing totally fake and 100 representing totally real news. \\ -- Do not say anything else about the result of your identification.}}
}
\end{center}

For each piece of news, we repeated this process 24 times to generate 24 distinct responses and the original AI advice set is shown in Figure~\ref{sequence_HMpro}. 

\vspace{1em}
\noindent \textbf{Human treatment.} The advice was randomly selected from another participant's first identification ($Initial\ Response$) within the same treatment.

\vspace{1em}
In each round, advice is independently randomized at the participant level, such that participants may receive different advice even when evaluating the same news article. This design helps mitigate potential biases arising from participants' beliefs about being observed by others, including the spotlight effect \citep{gilovich2000spotlight}.

\subsubsection{Second Identification}
\noindent In the final stage, participants submit a second identification ($Final\ Response$) of $HMpro$. Both the participant's initial response and the advice for that round are displayed on the slider using distinct colors (see Figures~I.7--I.10 in Online Appendix~I.1).

\subsection{Survey Questions About Prior Beliefs}

\noindent As well as demographic information-related questions (see Online Appendix~H), participants' prior beliefs were obtained using the following three questions. \label{surveyq}
\begin{enumerate}[
  label=\textbf{SQ\arabic*:},
  align=left,
  leftmargin=*,
  start=5
]
    \item \textit{Have you heard about ChatGPT?}
    \item \textit{How many days per week do you use ChatGPT on average?}
    \item \textit{In today's experiment, specifically in the task of ``assessing News' authenticity,'' which do you think can provide more accurate responses?}
\end{enumerate}

\subsection{Final Payoff}

\noindent The participants' final payoff consists of a fixed participation fee of 500 JPY and an additional performance--based payoff. Specifically, the additional payoff, $\pi$, was calculated based on the accuracy of one randomly selected identification from all their identifications throughout the experiment (a total of $30 \text{ rounds} \times 2\text{ identifications} = 60 \text{ identifications}$), determined using the following quadratic equation:
$$
\pi = \max\{0, \ 2300 - 0.3 \times (R_{rd} - HMpro^*_{rd})^2\} \ \text{JPY},
$$
where $R_{rd}$ is the randomly selected identification, and $HMpro^*_{rd}$ denotes the corresponding true $HMpro$ of the deepfake news in the selected round, after rounding.

\subsection{Materials and Summary}
\noindent The main experiment was conducted in the laboratory at the Institute of Social and Economic Research (ISER) at the University of Osaka on November 7 and 9, 2023, and October 28 and 29, 2024. We recruited 87 participants who were students at the University of Osaka registered in the ORSEE \citep{greiner2015subject} database of ISER. All participants were native Japanese speakers, 42 out of whom were assigned to the \textbf{Human} treatment and 45 to the \textbf{AI} treatment.\footnote{A power analysis (power$=0.8$,  Bonferroni-adjusted significant level $=0.025$) based on the result of a pilot experiment (the effect size,  $d=0.175$) suggests that we at least need  21 participants answering 30 tasks in each treatment.} In the final sample, 77\% of the participants were male, and 62\% were undergraduate students, predominantly from the following majors: 53\% engineering, 24\% medicine, 7\% law, and 4\% human science. Variable definitions are presented in Table~\ref{demodefi}, and comparisons of demographic data are illustrated in Figure~E.1 in Online Appendix~E.1.

During the experiment, participants were prohibited from using any of their own electronic devices, including smartphones and tablets. Although they completed the tasks on the laboratory's computers, Internet connectivity within the experiment software was also disabled.

After completing the 90-minute experiment, participants in the \textbf{Human} treatment earned an average final payoff of 2373 JPY, while those in the \textbf{AI} treatment earned 2429 JPY. As there were 30 round tasks for each participant, the final sample size was 2610 (1350 in the \textbf{AI} treatment and 1260 in the \textbf{Human} treatment). 

\label{prior}

\begin{table}[t] 
\centering
\begin{threeparttable}
\caption{Demographic Statistics}
\label{demodefi}
\small{
\renewcommand{\arraystretch}{1.5} 
\begin{tabular}{lm{8cm}ccccc} 
\toprule
Var.            & Definition & Min.  & Max.  & Avg. &S.D.\\
\midrule
age& Participants' age number.      & 18         & 44         & 22.8&3.34\\
freqGPT   & Average days per week using ChatGPT.     & 0    & 7   & 1.68&1.96\\
edulevel  & Participants' education level; = 1 if graduate; = 0 if undergraduate.   & 0  & 1 & 0.38&0.488\\
engr& Participants' major; = 1 if majoring in engineering.     & 0 & 1         & 0.53&0.502\\
male& Gender; = 1 if the participant is male.    & 0 & 1      & 0.77&0.423\\
progexp& Programming experience; = 1 if the participant has programming experience.   & 0  & 1 & 0.56&0.499\\
\bottomrule
\end{tabular}
}
\end{threeparttable}
\end{table}

\subsection{Hypotheses}

\noindent As discussed in the \textbf{Section}~\ref{section2}, deepfakes pose substantial societal risks, yet both human and machine-based detection---including AI detectors---remain imperfect and can generate secondary harms. At the same time, reliance on AI tools is increasingly observed across domains, but its implications remain context-dependent and not yet well understood. Despite these concerns, there is little systematic evidence on how individuals rely on AI-based advice when the task itself is to detect deepfake content. In such settings, excessive reliance on imperfect detection tools may fail to mitigate deepfake harms and may even introduce additional risks.

Against this backdrop, we begin by examining whether individuals rely more on AI-based advice than on advice from human peers when detecting deepfake news. This comparison provides a natural and testable benchmark for understanding the emergence of AI reliance in a context marked by growing societal concern over deepfakes.

\medskip
Accordingly, we propose the following hypothesis:
\begin{enumerate}[label=\textbf{H1:}]
    \item The reliance level on the external advice source is higher in the \textbf{AI} treatment than in the \textbf{Human} treatment.
\end{enumerate}

As noted, the news materials used in the tasks consist of three types of news---totally real, partially fake, and totally fake---with the proportion of AI-generated content ranging from 0\% to 100\%. This design reflects real-world deepfakes, which often combine AI-generated content with human editing rather than appearing as purely synthetic or purely human-made.

Prior work shows that ambiguous content is more difficult for individuals to evaluate than extreme cases \citep{friggeri2014rumor,lewandowsky2017beyond}. Consistent with this view, in the context of deepfake detection, estimating a specific proportion of human-written content ($HMpro \in [0,100]$) is substantially more challenging than making a binary judgment about whether an article is totally human-written or totally AI-generated ($HMpro \in \{0,100\}$). Consequently, when faced with news that contains a non-extreme proportion of AI-generated content, participants may experience greater uncertainty and become more inclined to rely on external advice.

This consideration leads to the following hypothesis:
\begin{enumerate}[label=\textbf{H2:}]
    \item The reliance level is higher when the news is partially fake than when it is totally fake or totally real.
\end{enumerate}

\section{Results of Main Experiment}
\label{section4}
\noindent This section presents the main findings of the experiment. We first examine participants' performance in detecting deepfake news. We then analyze their reliance on external advice and the corresponding treatment effects. Finally, we investigate whether the proportion of human-written content within each news article affects participants' performance and their reliance on advice.

\subsection{Performance}

\subsubsection{Basic Comparisons}

\begin{figure}[t]
\centering
\fbox{\includegraphics[scale=1]{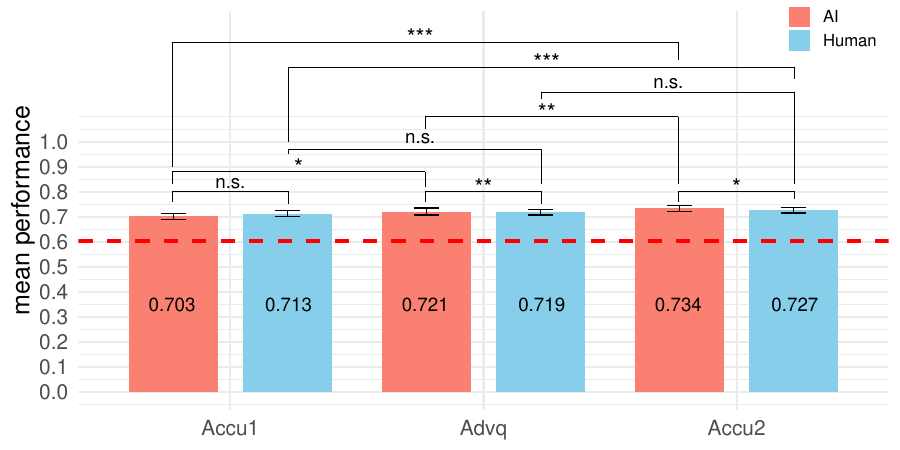}}
\caption{Overall Performance}\label{mainperformance}
\vspace{-0.8em}
\caption*{\small Note: $^+$ $p<0.1$, * $p<0.05$, ** $p<0.01$, *** $p<0.001$. ``n.s." means that the difference is not statistically significant at the 0.1 level. Error bars denote 95\% confidence intervals across participants. $Accu_1$, $Advq$, and $Accu_2$ are compared within treatments using the Wilcoxon signed-rank test, and between treatments using the Mann–Whitney U test. All reported p-values are adjusted for multiple comparisons using the Holm method. The red dashed line indicates the mean accuracy ($=0.603$) of an \textit{uninformative baseline}, which predicts a fixed value of 50 for every round.}
\end{figure}

\noindent We first investigated participants' performance in the deepfake detection task in our experimental setting by calculating their accuracy as

$$
Accu_{k,r}^i=1-\frac{|Response_{k,r}-HMpro^*_{r}|}{100}\text{,}$$ where $k\in\{1,2\}$, and $Accu_{1,r}^i$ and $Accu_{2,r}^i$ are participant $i$'s accuracy for ``$Initial\ Response$" ($Response_{1,r}^i$) and ``$Final\ Response$" ($Response_{2,r}^i$) in round $r$, respectively. $HMpro_r^*$ is the true $HMpro$ of the news article in round $r$.

Similarly, we defined the “advice quality” ($Advq$) for participant $i$ in round $r$ as the accuracy of the advice:
$$ 
Advq^i_r = 1 - \frac{|Advice^i_r-HMpro_r^*|}{100}
$$Importantly, although this measure is constructed using the true value $HMpro_r^*$, participants do not observe this benchmark. Thus, $Advq$ is interpreted as a researcher-side proxy for advice quality, which may shape participants' perceived advice quality.

Figure \ref{mainperformance} reports the mean $Accu_1$, $Accu_2$ and mean $Advq$. Participants' initial accuracy in detecting deepfake news---before receiving any advice---was 70.8\% on average, significantly above the \textit{uninformative baseline} (Wilcoxon signed-rank test, Holm-Adjusted $p<0.001$), indicating that the task is tractable and that participants were not responding randomly. The average quality of AI advice was 0.721, compared with 0.719 for advice from human peers. Although the difference in means is small, the Mann–Whitney U test shows a statistically significant difference ($\text{Holm-Adjusted }p = 0.0022$), suggesting that \textbf{AI can provide more accurate advice than Human peers}. 

Initial accuracy did not differ across treatments. However, final accuracy was significantly higher under the AI treatment than under the Human treatment (Holm-Adjusted $p = 0.0114$). As with advice quality, this difference appears to reflect a broader distributional shift: participants' post-advice accuracy tends to be higher when the advice comes from AI, even though the difference in mean accuracy remains modest.

OLS regression results are presented in Table A.1 in Online Appendix A. The coefficient on the treatment indicator $Tai$, which takes value 1 for AI treatment, is not statistically significant in most specifications. In contrast, the coefficient on $Advq$ is positive and statistically significant, indicating that the higher performance observed in the AI treatment is primarily driven by advice quality rather than the identity of the advice source.

\subsubsection{Performance Improvement}

\begin{figure}[t]
\centering
\fbox{\includegraphics[scale=0.7]{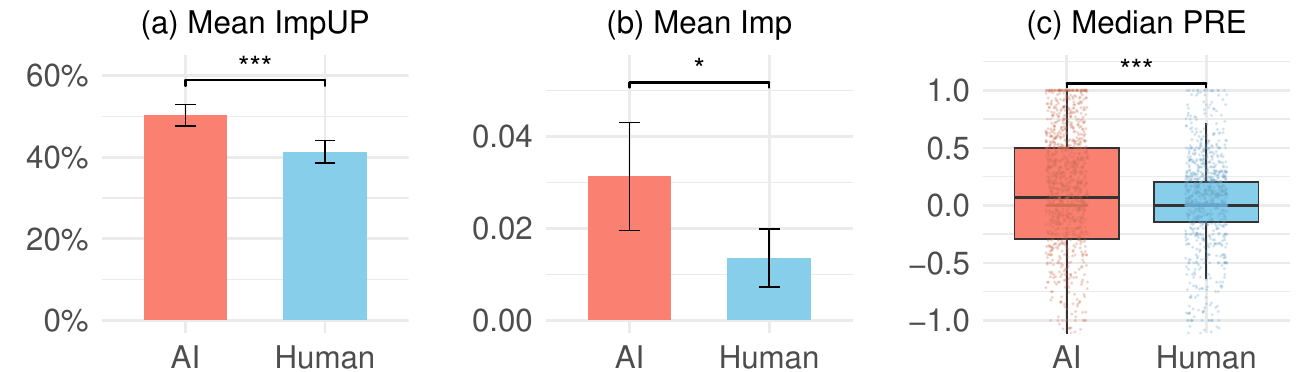}}
\caption{Performance Improvement}\label{performanceimp}
\vspace{-0.8em}
\caption*{\small Note: ImpUP is compared across treatments using Fisher's exact test, and Imp and PRE are compared across treatments using the Mann–Whitney U test. For PRE, 124 observations are excluded because participants with $Accu_1 = 1$ have undefined PRE values. $^+$ $p<0.1$, * $p<0.05$, ** $p<0.01$, *** $p<0.001$. ``n.s." means that the difference is not statistically significant at the 0.1 level. Error bars in (a) and (b) denote 95\% confidence
intervals across participants. In panel (c), each point corresponds to a participant–round PRE observation; boxes represent interquartile ranges with median lines, and whiskers extend to 1.5 times the interquartile range. The y-axis is restricted to the range $[-1,1]$ to highlight the central distribution.}
\end{figure}

\noindent Figure \ref{mainperformance} shows that, participants in both conditions improved after receiving external advice and revising their initial identifications, especially in AI treatment, which suggest that AI advice provides more effective support in detecting deepfake news than advice from human peers. We then plot the \textit{indicator for performance improvement}  ($\text{ImpUP}=1$ if $Accu_2 > Accu_1$, and $0$ otherwise), the \textit{raw performance improvement} ($\text{Imp} = Accu_2 - Accu_1$), and the \textit{proportional reduction in error} (PRE)\footnote{For round \(r\),
\( \mathrm{PRE}_r = \dfrac{\mathrm{Error}_{1,r}-\mathrm{Error}_{2,r}}{\mathrm{Error}_{1,r}}
= \dfrac{(1-Accu_{1,r})-(1-Accu_{2,r})}{1-Accu_{1,r}}
= \dfrac{Accu_{2,r}-Accu_{1,r}}{1-Accu_{1,r}}\in(-\infty,1]\), which measures the \emph{fraction of the initial error removed} by the second identification:
\(\mathrm{PRE}=1\) means the initial error is fully eliminated; \(\mathrm{PRE}=0\) means no change; \(\mathrm{PRE}<0\) indicates deterioration. PRE is computed when \(Accu_{1,r}<1\).} in Figure \ref{performanceimp}. 

Across all three measures, participants in the AI treatment perform better than those in the Human treatment: \textit{they are more likely to improve their accuracy, achieve larger improvements, and eliminate more initial errors after receiving AI advice}.

We further analyze these patterns by estimating Probit regressions for $\text{ImpUP}$ and OLS regressions for $\text{Imp}$. The results are reported in Tables A.2 and A.3 in Online Appendix A. 

Across both the Probit specification (ImpUP) and the OLS specification (Imp), advice quality ($Advq$) emerges as the strongest and most consistent predictor of performance improvement. \textbf{Higher-quality advice substantially increases both the likelihood and the magnitude of improvement}. On average, \textbf{AI advice generates slightly larger improvements than human advice}. When we include the interaction term $Tai \times Advq$, the pattern becomes clearer. The negative coefficient on $Tai$, combined with the large positive coefficient on the interaction term, indicates that \textbf{the effectiveness of AI advice is highly dependent on advice quality}. \textit{When advice quality is low, AI advice is less likely than human advice to produce performance gains. As advice quality increases, however, the relative effectiveness of AI advice rises rapidly.} \textbf{At sufficiently high levels of advice quality}, AI advice becomes substantially more effective than human advice in improving participants' performance. 

\begin{result}
\label{result1}
\textit{AI improves participants' performance primarily because it provides higher-quality advice.}
\end{result}

\subsection{Reliance Level}

\noindent The degree of ``reliance on external advice source" is measured by the ``weight of advice" \citep{onkal2009relative}, which is defined for participant $i$ in the task of round $r$, as follows.

$$\begin{aligned}WOA^i_r=\frac{Final\ Response^i_r - Initial\ Response^i_r}{Advice^i_r-Initial\ Response^i_r} \end{aligned}$$

This quantification approach produces a continuous outcome, where a higher WOA indicates a greater reliance on external advice. Additionally, a WOA greater than 0.5 indicates that the participant's final response is closer to the external advice than to their own initial response, representing a relatively higher degree of reliance on external advice. In contrast, a WOA less than 0.5 signifies a lower degree of reliance on external advice. 

In our analysis of the main experiment, we excluded 116 observations where WOA was undefined (i.e., when the initial response was equal to the advice given) and kept all negative values, resulting in an adjusted sample size of 2494 (1291 in the \textbf{AI} treatment and 1203 in the \textbf{Human} treatment).

\begin{figure}[t]
\centering
\fbox{\includegraphics[scale=0.75]{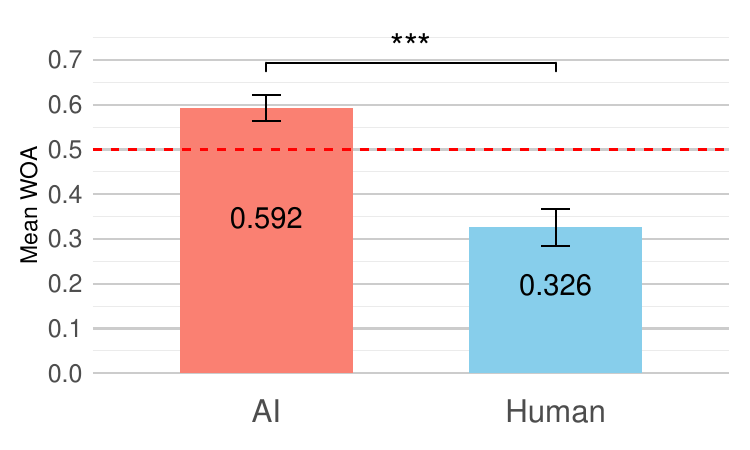}}
\caption{ WOA Across Treatments}\label{mainWOA}
\vspace{-0.8em}
\caption*{\small Note: $^+$ $p<0.1$, * $p<0.05$, ** $p<0.01$, *** $p<0.001$. ``n.s." means that the difference is not statistically significant at the 0.1 level. Error bars denote 95\%
confidence intervals across participants. WOA is compared across treatments using the Mann–Whitney U test. The red dashed line ($WOA = 0.5$) represents the baseline in which participants place equal weight on the advice and on their initial response.}
\end{figure}

Figure \ref{mainWOA} shows the mean WOA for the two treatments in the main experiment. The average WOA was 0.592 in the \textbf{AI} treatment and 0.326 in the \textbf{Human} treatment, with the former being significantly higher.
Both treatment means also differ significantly from the benchmark value of 0.5 (Mann–Whitney U test, $\text{Holm-Adjusted }p<0.001$).

These results indicate that participants in the \textbf{Human} treatment placed relatively little weight on advice from their peers, tending instead to adhere to their initial judgments. In contrast, participants in the \textbf{AI} treatment placed substantially more weight on advice from ChatGPT and were more willing to adjust their initial responses accordingly, providing strong support for \textbf{H1}.

\begin{result}
\label{result2}
\textit{People rely on ChatGPT more than their peers when detecting AI-generated content in deepfake news articles}
\end{result}

\subsection{The Role of Human-written Proportion}

\noindent In addition to the categorical news types described in Table~\ref{demodefi}, we examine whether variation in the proportion of human-written (non-AI-generated) content\footnote{Note that the proportion of human-written content is mechanically equivalent to the complement of the AI-generated proportion; higher values of one necessarily imply lower values of the other. We adopt the former throughout for consistency with the main text.} within each article influences participants' detection performance and their reliance on external advice, enabling a more fine-grained analysis of behavior in deepfake detection tasks.
\vspace{0.5em}

\noindent \textbf{Performance.} Tables A.1--A.3 in Online Appendix A report the regression results examining the role of human-written proportion in shaping performance outcomes. In Table A.1 (OLS regressions of performance), the coefficients on $isreal$ and $isfake$---indicators for totally human-written and totally AI-generated articles---are not statistically significant, and they remain insignificant in Table A.2 (Probit regressions of performance improvement). In contrast, both coefficients become significantly positive in Table A.3 (OLS regressions of performance improvement), indicating that participants improve more when evaluating extreme cases (totally human-written or totally AI-generated articles) than when evaluating mixed-content articles. The coefficient on $HMpro$ is significantly negative in Table A.1, although its magnitude is negligible, and it is not significant in Tables A.2 or A.3. This pattern suggests that although participants' initial performance decreases slightly as articles become more human-written, the proportion of human-written content does not systematically affect whether participants improve or by how much they improve after receiving advice.

\begin{result}
\label{result3}
\textit{The proportion of human-written content within a deepfake news article does not exert a systematic effect on participants' detection performance or on their improvement after receiving advice.}
\end{result}

\noindent \textbf{Reliance level.} Table A.4 in Online Appendix A reports the OLS regression results for reliance measured by WOA. Across all specifications, the coefficients on $HMpro$, $isreal$, $isfake$, and their interactions with the treatment indicator are not statistically significant. These findings indicate that the extent of AI generation in a deepfake news does not appear to influence reliance behavior in the deepfake detection task. Therefore, our second hypothesis, \textbf{H2}, is rejected.

\begin{result}
\label{result4}
\textit{Participants' reliance on external advice does not vary with the proportion of human-written content in the deepfake news article.}
\end{result}

\section{Additional Experiments}
\label{section5}
\subsection{New treatments and Hypotheses}
\noindent As a robustness check, we conducted three additional sessions on June 16, June 30, and October 8, 2025, in the laboratory of the ISER at the University of Osaka. Relative to the main experiment, these sessions introduced two new advice sources and moved the post–main-task survey to an earlier stage, positioned between the Quiz and Practice sections. The three additional treatments are summarized below.

\begin{enumerate}
    \item \textbf{Expert}. The survey was administered immediately after the Quiz and before the Practice stage. In the Main Task, participants received advice given by linguistic experts.\footnote{Participants were informed that each piece of advice came from an identification made by \textit{a linguistic expert}---professors, assistant professors, or lecturers specializing in linguistics. We collected Experts' advice via a separate questionnaire from 11 experts affiliated with the University of Osaka, Kwansei Gakuin University, and Keio University, who provided their own identification of at least five deepfake news articles used in the main experiment, without online search. Note that unlike human peer's initial evaluation, these experts were not given monetary incentive for providing accurate evaluation. For each of the 30 articles, we collected at least three expert identifications.}
    \item \textbf{preHuman}. The survey was administered before the Practice stage, same as in the \textbf{Expert} treatment instead of after the main task. In the Main Task, the advice shown to participants was a randomly selected first identification reported by \textit{participants in the \textbf{Human} condition of the main experiment reported above}.\footnote{Participants were informed that each piece of advice came from an first identification made by a participant in the experiment conducted in the previous year.}
    \item \textbf{AIadd}. The survey was administered immediately after the Quiz and before the Practice stage. All other procedures replicated the \textbf{AI} treatment in the main experiment.
\end{enumerate}

The purpose of the \textbf{Expert} treatment is to extend the comparison beyond ChatGPT and student peers by introducing advice generated by linguistic experts. The \textbf{preHuman} treatment is designed to examine whether the reliance on advice from human peers in the main experiment may have been affected by social interactions or other-regarding preferences.\footnote{We thank anonymous reviewer for pointing out this possibility.} The \textbf{AIadd} treatment is intended to test whether the timing of the survey influences participants' prior beliefs about the advice source.\footnote{The \textbf{Expert} and \textbf{preHuman} treatments were preregistered at aspredicted.org (\#233371). The \textbf{AIadd} treatment was not preregistered because the main task remained unchanged.} 

Previous research suggests that people tend to trust responses generated by LLMs more than those provided by human experts \citep{shekar2025people}. We also assume that participants generally do not distinguish between advice coming from peers in the same session and peers who participated earlier. Therefore, for the two newly added treatments, we have the following hypotheses:

\begin{enumerate}[label=\textbf{H3:}]
    \item Compared to linguistic experts, people tend to rely more on ChatGPT to detect deepfake news.
\end{enumerate}

\begin{enumerate}[label=\textbf{H4:}]
    \item Compared to human peers (students who participated in this experiment before), people tend to rely more on ChatGPT to detect deepfake news.
\end{enumerate}

\subsection{Results of Additional Experiment}

\noindent We recruited 133 participants who were also students at the University of Osaka registered in the ORSEE \citep{greiner2015subject} database of ISER. All participants were native Japanese speakers, 44 out of whom were assigned to the \textbf{Expert} treatment, 47 to the \textbf{preHuman} treatment and 42 to the \textbf{AIadd} treatment. Participants in the \textbf{Expert} treatment earned an average final payoff of 2420 JPY, while those in the \textbf{preHuman} treatment earned 2425 JPY, and those in the \textbf{AIadd} treatment earned 2513 JPY. Comparisons of demographic data are illustrated in Figure E.1 in Online Appendix E.1. 

\subsubsection{Advice Quality}
\label{performanceadvq}

\begin{figure}[t]
\centering
\fbox{\includegraphics[scale=0.67]{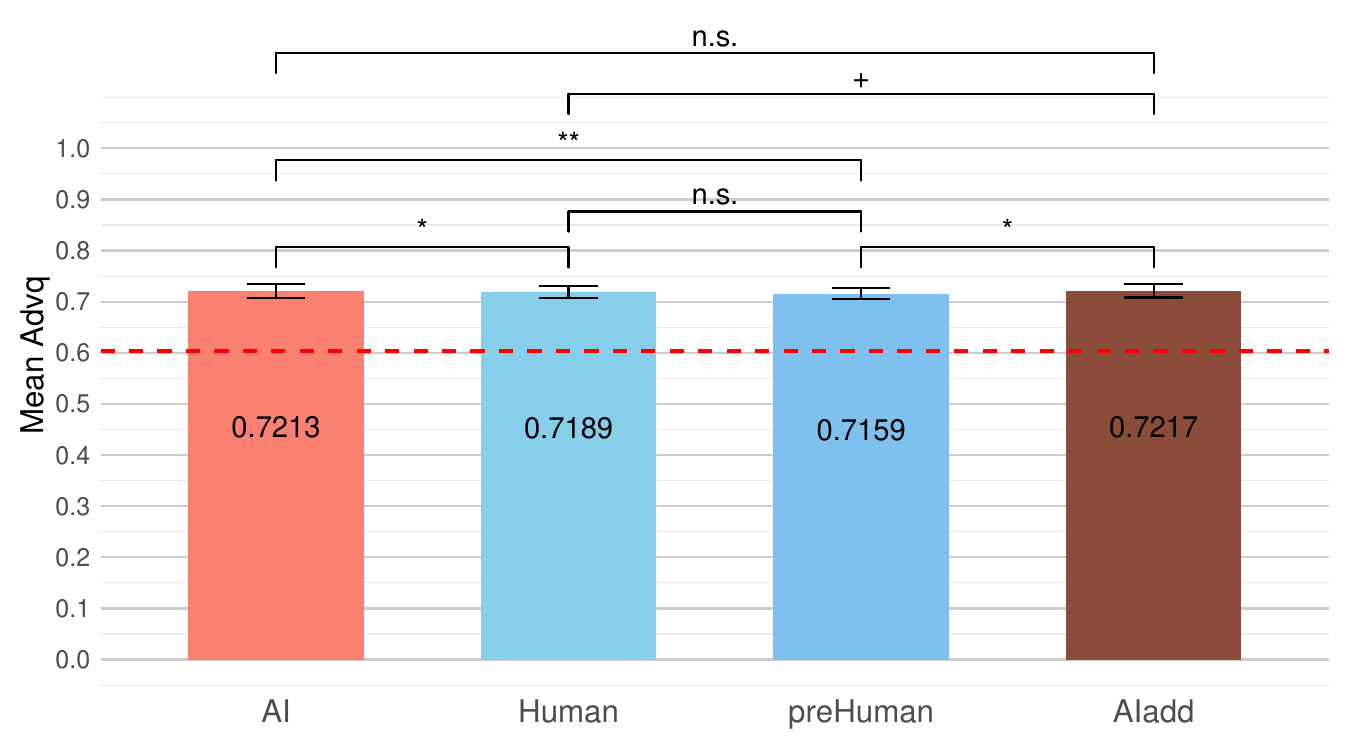}}
\caption{Mean $Advq$ Across Treatments (Excluding \textbf{Expert})}\label{addAdvq}
\vspace{-0.8em}
\caption*{\small Note: $^+$ $p<0.1$, * $p<0.05$, ** $p<0.01$, *** $p<0.001$. ``n.s." means that the difference is not statistically significant at the 0.1 level. Error bars denote 95\% confidence intervals across participants. The Expert treatment is excluded from this figure because expert advice was collected via an unincentivized questionnaire and is therefore not comparable in advice quality to other treatments. $Advq$ is compared across treatments using the Mann–Whitney U test. All reported p-values are adjusted for multiple comparisons using the Holm method. The red dashed line indicates the mean accuracy ($=0.603$) of an \textit{uninformative baseline}, which predicts a fixed value of 50 for every round.}
\end{figure}

\noindent Figure~\ref{addAdvq} reports the quality of advice shown to participants across \textbf{AI}, \textbf{Human}, \textbf{preHuman} and \textbf{AIadd} treatments, which shows that
ChatGPT achieves higher accuracy than human participants in the deepfake detection task, and therefore can provide higher-quality advice. 

Here, we exclude the \textbf{Expert} treatment from $Advq$ comparisons because experts' advice was collected under a very different condition from human advice (in particular, it was not incentivized for accurate evaluation); therefore, we refrain from presenting and comparing their accuracy with other advice.\footnote{For reference only, the mean value of $Advq$ for expert advice is 0.6919.} Nevertheless, expert advice's $Advq$ is included in subsequent analyses that examine the relationship between $Advq$ and participants' reliance and detection performance. This is because participants were not informed about the incentive structure behind expert advice, and thus any differences in incentive provision do not affect participants' beliefs or decision-making conditional on the realized $Advq$.

\subsubsection{Performance}

\begin{figure}[t]
\centering
\fbox{\includegraphics[scale=0.67]{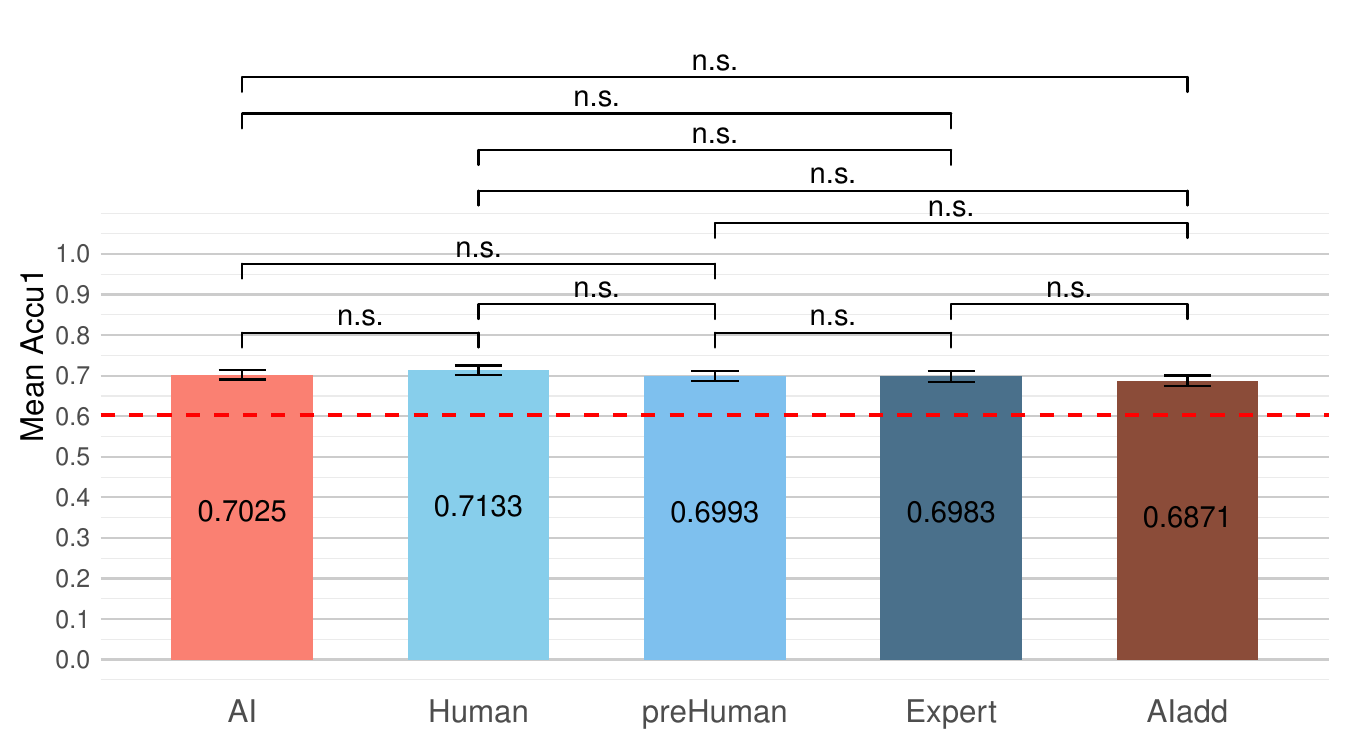}}
\caption{Mean $Accu_1$ Across Treatments} \label{accu1add}
\vspace{-0.8em}
\caption*{\small Note: $^+$ $p<0.1$, * $p<0.05$, ** $p<0.01$, *** $p<0.001$. ``n.s." means that the difference is not statistically significant at the 0.1 level. Error bars denote 95\%
confidence intervals across participants. $Accu_1$ is compared across treatments using the Mann–Whitney U test. All reported p-values are adjusted for multiple comparisons using the Holm method.}
\end{figure}

\noindent \textbf{Initial Detection Accuracy.} Figure \ref{accu1add} shows that $Accu_1$ does not differ systematically across treatments. Columns (1)--(3) of Table B.1 in Online Appendix B further indicate that the human-written proportion itself has no meaningful effect on baseline detection accuracy. The indicators for extreme cases---articles that are totally human-written ($isreal$) or totally AI-generated  ($isfake$)---are statistically significant, but their magnitudes are small: the coefficients imply only a 3\%--4\% point reduction in accuracy. This pattern suggests that participants behave somewhat cautiously---and may hedge against potential errors---when facing boundary cases in the absence of advice, though the practical impact on baseline performance remains limited.
\vspace{0.5em}

\begin{figure}[t]
\centering
\fbox{\includegraphics[scale=0.67]{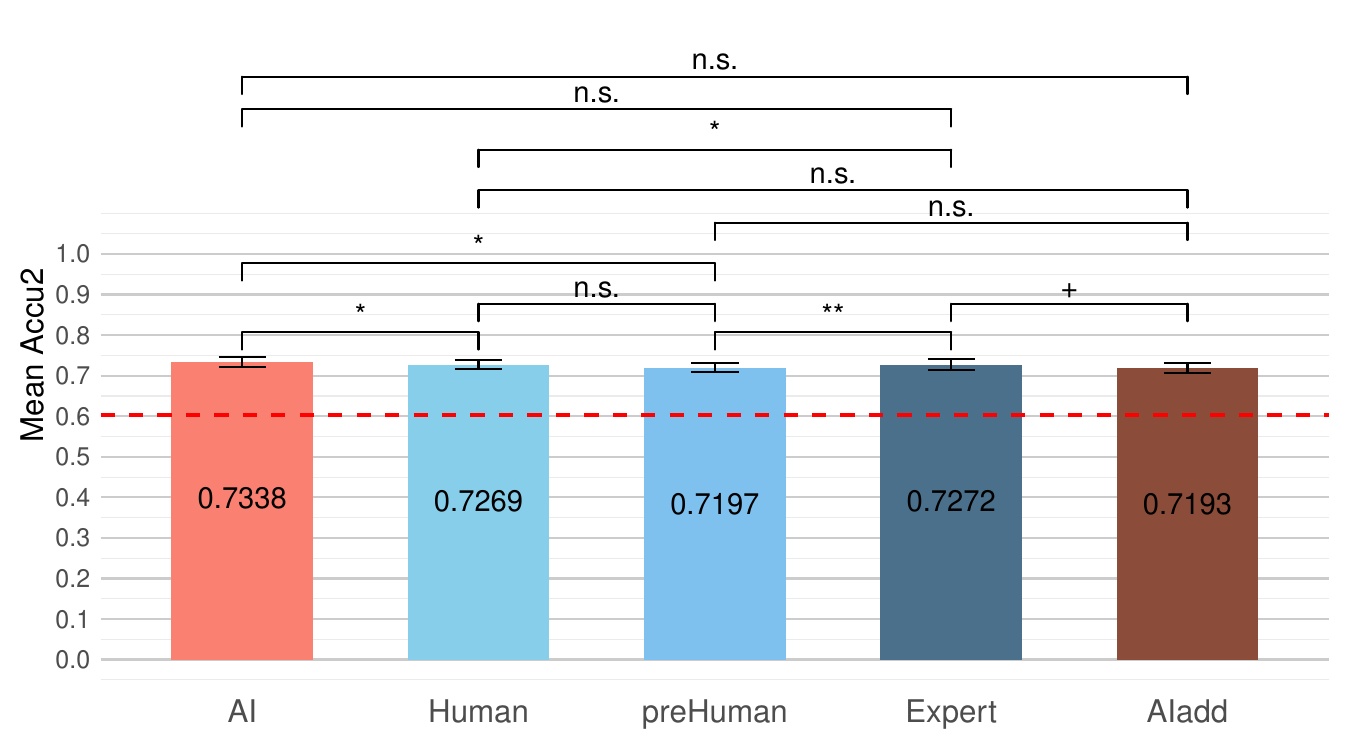}}
\caption{Mean $Accu_2$ Across Treatments}\label{accu2add}
\vspace{-0.8em}
\caption*{\small Note: $^+$ $p<0.1$, * $p<0.05$, ** $p<0.01$, *** $p<0.001$. ``n.s." means that the difference is not statistically significant at the 0.1 level. Error bars denote 95\% confidence intervals across participants. $Accu_2$ is compared across treatments using the Mann–Whitney U test. All reported p-values are adjusted for multiple comparisons using the Holm method. The red dashed line indicates the mean accuracy ($=0.603$) of an \textit{uninformative baseline}, which predicts a fixed value of 50 for every round.}
\end{figure}

\noindent \textbf{Final Detection Accuracy.} Figure \ref{accu2add} reports the average $Accu_2$ across the five treatments. The \textbf{AI} treatment achieves the highest final accuracy. The \textbf{Expert} treatment also attains a higher $Accu_2$ than both the \textbf{Human} and \textbf{preHuman} treatments. The difference between the \textbf{AI} and \textbf{Expert} treatments is not statistically significant.

Columns (4)--(6) of Table B.1 and the corresponding interaction results in Table B.2 in the Online Appendix B show that the human-written proportion ($HMpro$) has a statistically significant but substantively negligible effect on $Accu_2$: although the coefficient is negative and significant, its magnitude (about –0.0003) implies an effect that is economically minimal.

By contrast, advice quality ($Advq$) plays a dominant role. A one percentage point increase in advice quality (i.e., a 0.01 increase on the 0–1 scale) raises final accuracy by roughly 0.0055 in the baseline specification. This effect is even stronger in the AI treatment: the interaction term (0.223) implies that the total marginal effect of advice quality for AI advice is approximately $0.499 + 0.223 = 0.722$, highlighting the particularly high returns to advice quality when the advice comes from ChatGPT.

\vspace{0.5em}
\noindent \textbf{Performance Improvement.} Consistent with the results from the main experiment, participants in the three additional sessions also improved their performance after receiving advice, with $Accu_2$ being significantly higher than $Accu_1$ (Wilcoxon signed-rank test, $p<0.0001$). 

Figures B.1 and B.2 in Online Appendix B further report the average Imp and ImpUP across treatments. While no statistically significant differences are observed in raw performance improvement across treatments, participants who receive advice from \textbf{AI} or \textbf{Expert} exhibit a higher probability of performance improvement than those who receive advice from peer sources.

Tables B.3 and B.4 in Online Appendix B report the corresponding regression analyses for performance improvement. Across all specifications, the human-written proportion has no significant effect on the likelihood of improvement, and its effect on raw improvement---while sometimes statistically significant---is economically negligible. By contrast, advice quality ($Advq$)---particularly AI advice---exhibits a large, stable, and highly significant positive effect in every specification. Its magnitude exceeds that of treatment indicators, the human-written proportion, and the extreme-news indicators ($isfake$, $isreal$) by an order of magnitude.

This pattern indicates that performance differences across treatments are driven overwhelmingly by differences in the quality of advice participants receive, rather than by treatment identity per se. Specifically, AI advice tends to generate larger improvements because it systematically provides higher-quality guidance. Variation in $Accu_2$ and improvement therefore arises primarily because treatments expose participants to advice pools of differing quality. This supports our conclusion in Result~\ref{result1}: \textit{treatment differences in the additional experiment do not operate independently of advice quality; rather, advice quality is the dominant channel through which these differences manifest}.

\subsubsection{Reliance Level}

\begin{figure}[t]
\centering
\fbox{\includegraphics[scale=0.67]{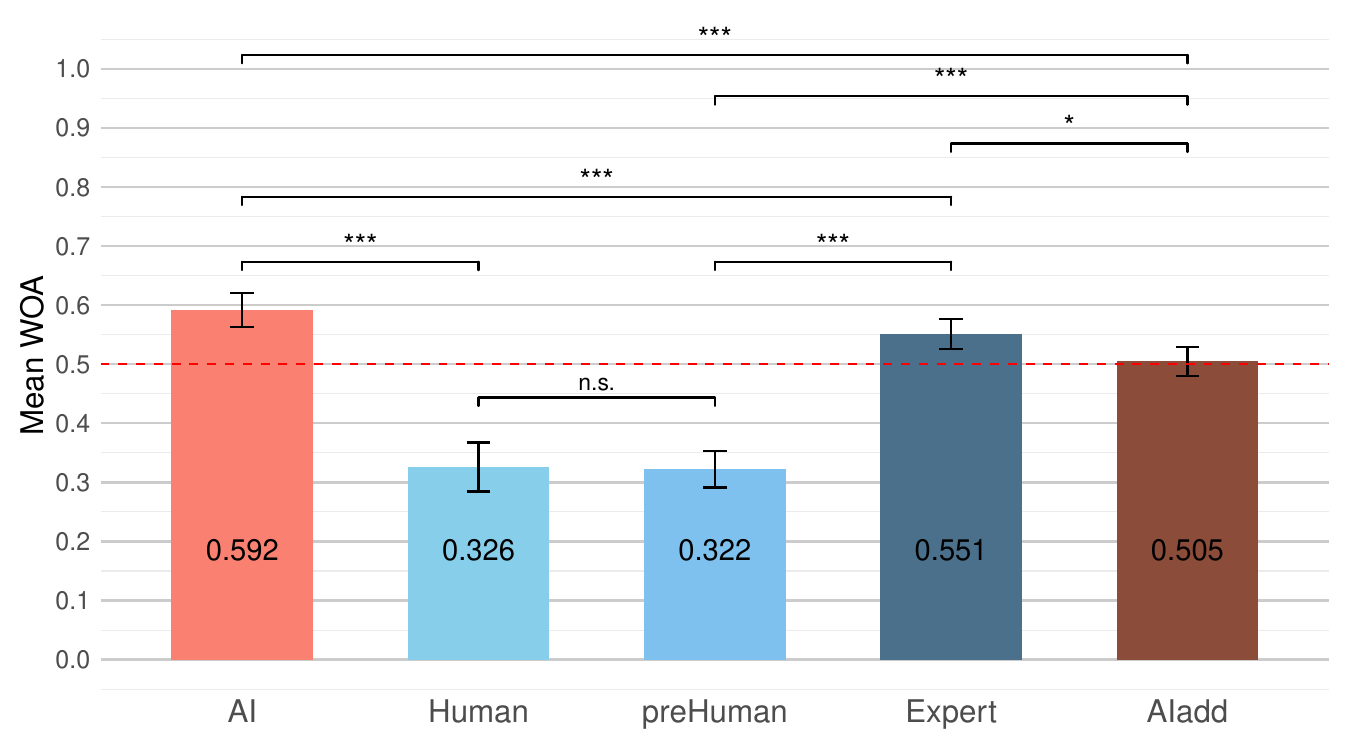}}
\caption{ WOA Across Treatments}\label{woaadd}
\vspace{-0.8em}
\caption*{\small Note: $^+$ $p<0.1$, * $p<0.05$, ** $p<0.01$, *** $p<0.001$. ``n.s." means that the difference is not statistically significant at the 0.1 level. Error bars denote 95\%
confidence intervals across participants. WOA is compared across treatments using the Mann–Whitney U test. All reported p-values are adjusted for multiple comparisons using the Holm method. The red dashed line ($WOA = 0.5$) represents the baseline in which participants place equal weight on the advice and on their initial response.}
\end{figure}

\noindent Figure~\ref{woaadd} reports the WOA values across the five treatments. The ordering of reliance is $WOA_{AI}>WOA_{Expert}>WOA_{AIadd}>WOA_{Human}\approx WOA_{preHuman}$. The comparisons between \textbf{Human} and \textbf{preHuman}, and between \textbf{AI / AIadd} and \textbf{Human / preHuman}, confirm \textbf{H4}: \textit{reliance on peer advice in the main experiment is not driven by social interaction or other-regarding preferences. Participants rely substantially more on ChatGPT than on human peers}.

A more nuanced pattern emerges when comparing reliance across the three non-peer treatments. Although reliance in the \textbf{AI} treatment is significantly higher than in the \textbf{Expert} treatment (Holm-adjusted $p=0.00002$), this comparison spans different implementation years (2023--2024 vs. 2025) and thus may reflect temporal or contextual differences rather than source credibility. By contrast, the \textbf{Expert} and \textbf{AIadd} treatments were both conducted in 2025 under an identical procedural flow, providing a cleaner within-year comparison of advice sources.\label{WOAadd}\footnote{A detailed discussion of this cross-wave divergence is provided in Online Appendix C.} Reliance is significantly higher for \textbf{Expert} than for \textbf{AIadd} (Holm-adjusted $p=0.01308$), indicating that when procedural and temporal factors are held constant, participants place more weight on advice attributed to linguistic experts than on advice attributed to ChatGPT. Accordingly, \textbf{H3} is not supported.

\begin{result}
\label{result5}
\textit{Participants assign the greatest reliance to advice from linguistic experts, followed by advice from ChatGPT, and rely the least on advice from human peers when detecting deepfake news.}
\end{result}

Table B.5--B.7 in Online Appendix B report the OLS regression results for participants' reliance behavior with respect to the human-written proportion in deepfake news. Across almost all specifications, the coefficients on $HMpro$, $isreal$, $isfake$, as well as their interactions with the treatment indicators, are statistically insignificant. These results further reinforce our earlier conclusion in Result~\ref{result4} that the human-written proportion of the news content does not meaningfully affect participants' reliance level.\vspace{1em}

\section{Discussions}
\label{section6}
\subsection{Prior Beliefs}

\noindent We elicit participants' prior beliefs using three survey questions reported in Section~\ref{surveyq}. Since all participants answered ``yes'' to \textbf{SQ5}, we focus on responses to \textbf{SQ6} and \textbf{SQ7}.

\textbf{SQ6} measures the self--reported frequency of using ChatGPT ($freqGPT$). Figure~E.1 in Online Appendix~E.1 reports the average values of $freqGPT$ across treatments.\footnote{The $freqGPT$ levels in the additional experiment are substantially higher than those in the main experiment. A likely explanation is that the main experiment was conducted in 2023--2024, whereas the additional experiment took place in 2025, a period during which public familiarity with and usage of AI tools increased markedly.}

\textbf{SQ7} elicits participants' beliefs about whether GAI or humans would perform better in the deepfake detection task. Importantly, this question was asked before participants observed any task outcomes or payoff information. In the two treatments of the main experiment, this question was asked after the main task but before the display of results and payoffs. In the three treatments of the additional experiment, the same question was asked before the main task began.

Based on responses to \textbf{SQ7}, we constructed a dummy variable, $prefAdvSrc$, capturing ``participants' relative preference for the offered advice source,” or ``prior beliefs about relative advice effectiveness", coded as:$$
prefAdvSrc_i =
\begin{cases}
1, & \text{if $i$ in AI/AIadd treatment} \\
   & \quad \text{and trusted AI performs better}, \\
1, & \text{if $i$ in Human/preHuman/Expert treatment} \\
   & \quad \text{and trusted Human performs better}, \\
0, & \text{otherwise}.
\end{cases}
$$Therefore, a $prefAdvSrc_i=1$ indicates that participant $i$ received advice from the source they prefer, and $prefAdvSrc_i=0$ indicates that they believe the advice source may not offer them good advice. 

\begin{figure}[t]
\centering
\fbox{\includegraphics[scale=0.67]{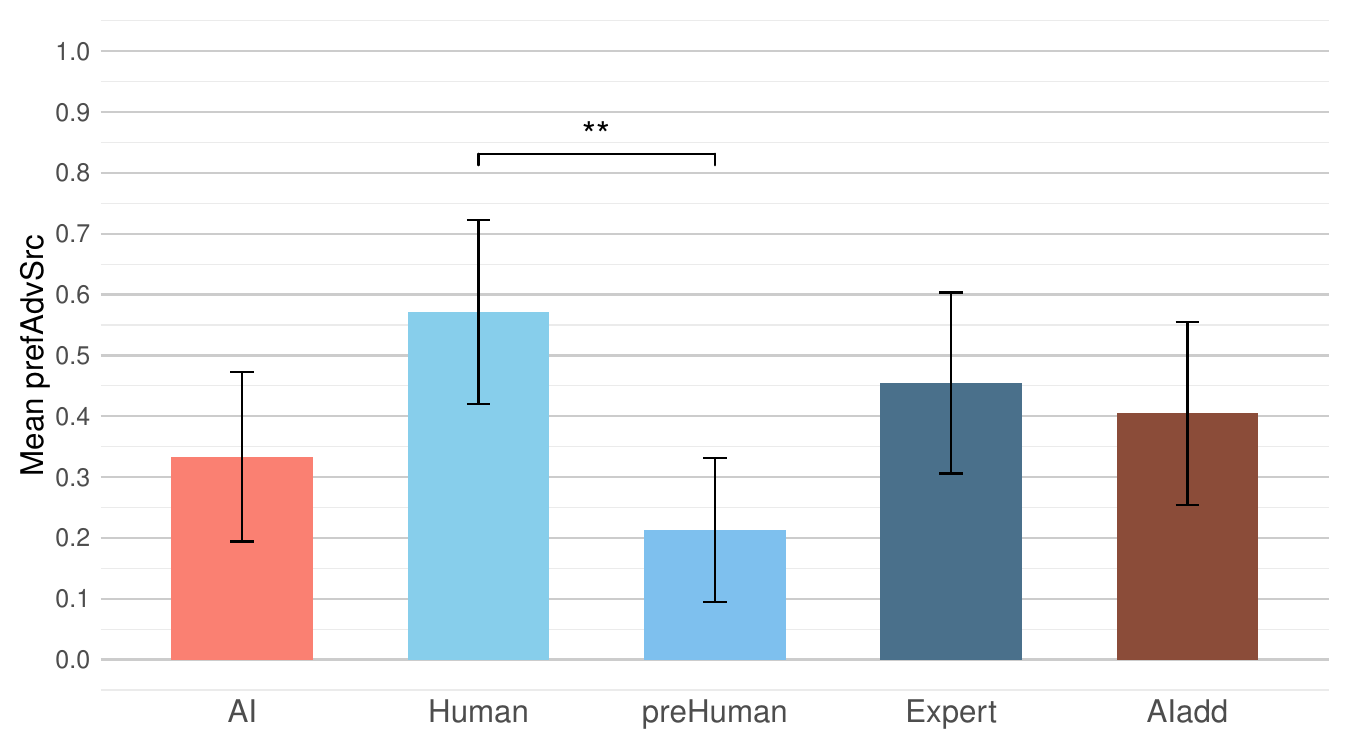}}
\caption{Mean $prefAdvSrc$ Across Treatments}\label{prefAdvSrc}
\vspace{-0.8em}
\caption*{\small Note: $^{+} p<0.1$, $^{*} p<0.05$, $^{**} p<0.01$, $^{***} p<0.001$. 
``n.s.'' indicates that the difference is not statistically significant at the 10\% level. 
Error bars denote 95\% confidence intervals across participants. 
Pairwise comparisons across treatments are conducted using Fisher's exact tests. 
All reported p-values are adjusted for multiple comparisons using the Holm method. None of the pairwise differences are statistically significant except for the comparison between the Human and preHuman treatments ($\text{Holm Adjusted } p=0.0091$).}
\end{figure}

Figure \ref{prefAdvSrc} presents the $prefAdvSrc$ across treatments. Among all pairwise comparisons, only the difference between the \textbf{Human} and \textbf{preHuman} treatments is statistically significant; all other comparisons show no significant differences. In particular, we do not detect a statistically significant difference in $prefAdvSrc$ between the \textbf{AI} (fielded in 2023/2024) and \textbf{AIadd} (fielded in 2025) samples. However, because questionnaire timing and calendar time vary simultaneously across these experimental, this comparison does not cleanly identify a pure timing effect on prior beliefs; rather, it suggests that any timing- or cohort-driven differences in stated priors are not large enough to be statistically detected in our data.

Given this pattern, it is reasonable to pool observations from the main and additional experiments when analyzing $prefAdvSrc$. OLS regression results are reported in Table B.8 in Online Appendix B. The coefficient on $prefAdvSrc$ is positive and statistically significant, indicating that \textit{participants are more likely to rely on advice received from a source they prefer}. In contrast, frequent use of ChatGPT in daily life does not translate into greater reliance on AI advice in the experimental setting. In particular, participants who received ChatGPT's advice and believed that GAI outperforms humans exhibited higher reliance on AI advice. Moreover, we find little evidence that differences in the preference-elicitation protocol materially alter stated preferences or their association with reliance. Specifically, the interaction patterns between $prefAdvSrc$ and the AI-based treatments are qualitatively similar in \textbf{AI} and \textbf{AIadd}. This stability supports pooling the main and additional experiments for the purpose of analyzing how $prefAdvSrc$ relates to reliance behavior, while acknowledging that calendar time and elicitation timing vary jointly across experimental waves.

\subsection{The Joint Role of Reliance and Advice Quality}

\noindent To better interpret the performance results, we examine how reliance and advice quality jointly shape performance improvement. In Section \ref{performanceadvq}, we show that advice quality ($Advq$) has a large and stable association with participants' performance, substantially stronger than the treatment indicators capturing advice source. The significant interaction between $Tai$ and $Advq$ in Tables B.2 and B.4 in Online Appendix B further suggests that, conditional on advice quality, participants receiving AI advice are more likely to translate advice into performance improvements than those receiving peer advice.

\begin{figure}[t]
\centering
\fbox{\includegraphics[scale=0.67]{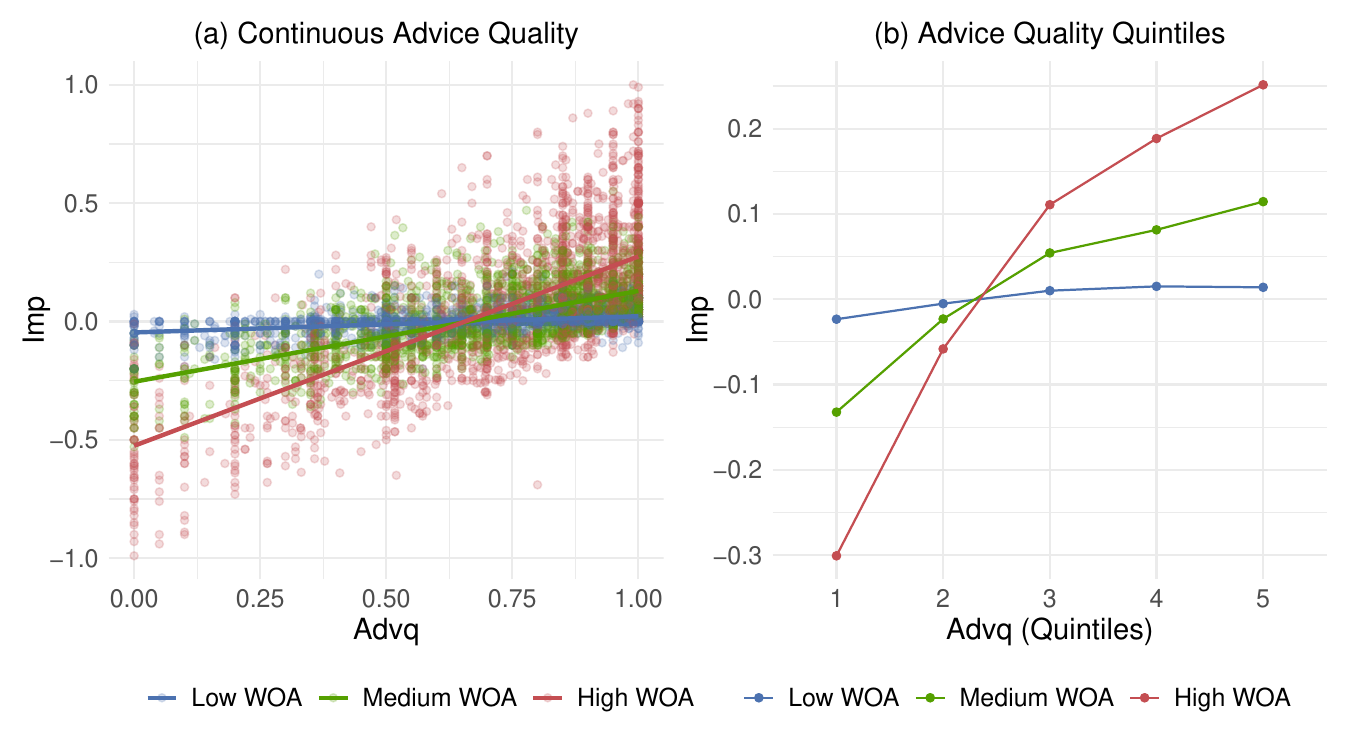}}
\caption{Advice Quality, WOA \& Performance Improvement}\label{advice_advq}
\vspace{-0.8em}
\caption*{\small Note: All observations are classified into low, medium, and high WOA groups based on terciles of the WOA distribution. Panel (a) plots the relationship between continuous advice quality and performance improvement, with fitted linear trends shown separately for each WOA group. Panel (b) provides a non-parametric illustration by grouping advice quality into quintiles and plotting mean performance improvement within each bin for the three WOA groups. 
}
\end{figure}

These findings motivate a simple distinction: \textit{advice source mainly shapes reliance behavior (WOA)---the weight participants place on advice---whereas advice quality determines how beneficial advice is conditional on being taken}. Figure \ref{advice_advq} visualizes this mechanism. The relationship between advice quality and performance improvement is weak among participants with low WOA, but becomes progressively stronger for those with medium and high WOA, with the steepest gradient among the high-WOA group.\footnote{An OLS regression of Imp on $Advq$, WOA, and $Advq\times \text{WOA}$ yields a positive and significant interaction ($\beta=0.513$, $p<0.001$; participant-clustered standard errors).}

Importantly, participants do not observe objective advice quality at the time of decision making. Any association between ex post advice quality and their behavior must therefore operate through participants' perceived quality when evaluating the advice. Thus, the pattern in Figure \ref{advice_advq} implies that ex post advice quality translates into performance improvements primarily when participants choose to rely on the advice, consistent with a gating role of reliance.

At the same time, high reliance should not be interpreted as normative superiority of a given advice source. In this sense, \textbf{reliance governs whether advice can affect performance, while advice quality determines whether such reliance is ultimately beneficial}.

Supporting this interpretation, Tables B.9 and B.10 in the Online Appendix B show that higher ex post advice quality\footnote{Measured both contemporaneously ($Advq$) and with a one-period lag ($AdvqLag$).} is positively correlated with reliance behavior. \textbf{Although participants do not observe objective advice quality at the time of decision making, this correlation suggests that their subjective evaluation of advice quality is, on average, aligned with realized advice quality.} We interpret this correlation as suggesting that objective advice quality is positively associated with participants' perceived advice quality when evaluating the advice.

\subsection{Decomposition of Reliance}

\noindent Our main analysis relies on the WOA to quantify reliance. Although widely used, WOA has a well-known limitation: when a participant's initial response coincides with the advice, the denominator becomes zero, producing undefined or extreme values that obscure the underlying behavioral process. Rather than clipping these outliers, we exclude only the undefined cases (4.4\% of observations). However, this approach still prevents us from using the full sample and ignores some extreme but behaviorally meaningful instances of reliance.

To address this issue---and to provide both a robustness check for the WOA-based analysis and a deeper examination of the advice-taking mechanism---we adopt the activation–integration framework of \citet{vodrahalli2022humans}, which conceptualizes reliance as a two-stage process: whether participants become activated to use the advice, and, conditional on activation, how strongly the advice is integrated into the final judgment. We implement this framework using a Heckman selection model \citep{heckman1974shadow, heckman1979sample}.

The detailed specification and estimation results are reported in Online Appendix D; here we summarize the main findings.

\textbf{Activation}---\textit{whether participants choose to take the advice}---is strongly predicted by the advice source, prior beliefs, the advice–initial gap (gap between the advice and initial response), and experienced lagged advice quality: participants are significantly more likely to take AI or Expert advice, more likely to be activated when they believe the source is effective, and more likely to be activated after both a larger advice–initial gap and recent positive experience with advice quality.

Conditional on activation, \textbf{Integration}---\textit{the extent to which participants move toward the advice}---reflects a different set of forces. Integration is shaped by the advice source and prior beliefs, but, unlike activation, it is negatively affected by the advice–initial gap. This pattern suggests that large gaps may draw participants into considering the advice but subsequently make them more cautious about how far to adjust. Consistent with this, experienced lagged advice quality does not significantly predict integration without Heckman's correction.

\subsection{Other Robustness Checks}

\noindent Additional robustness checks are reported in Online Appendix E. Here we briefly summarize the main findings.
\vspace{1em}

\noindent \textbf{Demographics.} We examine heterogeneity with respect to demographic characteristics reported in Table \ref{demodefi}. The results show no systematic differences in either reliance or performance across demographic groups.

\vspace{1em}
\noindent \textbf{Decision Time.} We further investigate whether decision time influences advice reliance or performance, using measures of the time spent on the first identification, the second identification, and the average reading time per character. We find no evidence of meaningful heterogeneity along these dimensions.
\vspace{1em}

\noindent \textbf{News Categories.} We classify news articles based on topic and content into four initial categories and then aggregate them into two broader groups for balance: \textbf{HARD} news (fact-oriented and policy-relevant domains) and \textbf{SOFT} news (lifestyle, cultural, sports, and entertainment topics). Whether a deepfake belongs to the \textbf{HARD} category has no significant effect on reliance. However, participants find \textbf{HARD} deepfake news more difficult to detect. Relative to human advice, AI advice helps mitigate this difficulty, improving detection performance in \textbf{HARD} domains.
\vspace{1em}

\noindent \textbf{Learning Effects.} Using both regression-based time controls and block-level comparisons, we find no evidence of learning in advice reliance. In contrast, performance improves over rounds as participants gain experience in detecting deepfake news. This learning effect is particularly pronounced among participants receiving AI advice, whose performance improvements are larger in the final ten rounds.

\section{Conclusions}
\label{section7}
\noindent This paper studies how individuals rely on different advice sources when detecting deepfake news. We implement a laboratory deepfake-detection task in which participants identify the proportion of human-written content in synthetic news articles and receive advice from ChatGPT (GPT-4), human peers, or linguistic experts. Reliance is measured behaviorally using the WOA, and performance is evaluated by accuracy and performance improvement.

Four main findings emerge. First, participants rely more on GPT-4 than on human peers when detecting GPT-2–generated deepfake news. This difference is robust across experimental waves and persists even when peer advice is drawn from earlier sessions. Second, the human-written proportion within an article---and equivalently, the extent of AI generation---does not systematically affect performance or reliance behavior. Third, advice from ChatGPT improves performance relative to peer advice, and this advantage is explained primarily by advice quality rather than by the AI label per se: higher-quality advice generates larger performance improvements regardless of source, and the AI treatment performs better largely because it exposes participants to a higher-quality advice pool. Fourth, the relative reliance on experts versus ChatGPT is mixed across waves: Reliance on ChatGPT exceeds reliance on experts in the main experiment, whereas reliance on experts exceeds reliance on ChatGPT when both are implemented under the same additional procedural flow in 2025. 

In addition to documenting treatment differences in performance and reliance, our analysis shows that participants' prior beliefs systematically shape their reliance decisions. Consistent with the activation–integration framework of \citet{vodrahalli2022humans}, our results further indicate that reliance unfolds sequentially, with different factors governing whether advice is taken and how strongly it is integrated. We also document heterogeneity with respect to news categories and learning effects in performance, while finding no corresponding learning effects in advice reliance.

Overall, our findings extend the study of ``AI reliance" to the domain of deepfake detection and highlight the dual role of GAI---as both a potential source of misinformation and a tool for mitigating it. In settings where the object of detection is AI-generated content, reliance on AI-based advice is not necessarily detrimental and can improve performance when advice quality is sufficiently high. Importantly, this should not be interpreted as implying that ChatGPT is relied upon more than any human source. Rather, relative reliance depends on the type of advisor rather than a simple distinction between AI and human advice. From a policy perspective, the effectiveness of AI-based detection tools depends jointly on their objective quality and on public beliefs about who---or what---is trustworthy.

Several limitations point to directions for future research. First, our deepfake materials were generated using GPT-2. As frontier models continue to produce increasingly human-like content, both detection difficulty and reliance patterns may evolve, raising the need to test with more advanced generators. Second, we employ prompted GPT-4 as the AI-based detection tool, whereas real-world users often interact with dedicated detection systems that differ in interface design, transparency, and feedback structure. Future studies could compare reliance on conversational AI with reliance on specialized detectors and examine how reliance varies across models. Third, our study is conducted in a laboratory environment. While this setting affords strict control---such as preventing participants from consulting external AI tools---it also limits sample size due to cost constraints. Field or online implementations with stronger scalability, but still with appropriate control over participants' information sources, would be a valuable complement. Fourth, we do not directly observe the strategies participants use to detect deepfake news. Because deepfake news may involve both factual inconsistencies and stylistic cues associated with AI-generated text, future work could examine the relative roles of these dimensions and how individuals trade off between them in detection tasks.\footnote{Since the completion of this study, evidence from a related study \citep{FuHanaki2025Style} indicates that 88.6\% and 82.9\% of participants reported relying on linguistic and contextual cues, respectively, whereas only 27.2\% reported relying on factual verification when detecting AI-generated content.}

\newpage
\bibliography{Ref/Ref2404,Ref/Ref2410,Ref/Ref2511}
\newpage

\end{document}